# A Tutorial of the Mobile Multimedia Wireless Sensor Network OMNeT++ Framework

Zhongliang Zhao[*], Denis Rosário[*,†], Torsten Braun[*], Eduardo Cerqueira[†]
[*]Institute of Computer Science and Applied Mathematics, University of Bern, Switzerland
[†]Faculty of Computer Engineering and Telecommunication, Federal University of Para, Brazil
Email:zhao@iam.unibe.ch, denis@ufpa.br, braun@iam.unibe.ch, cerqueira@ufpa.br

*Abstract*—In this work, we will give a detailed tutorial instruction about how to use the Mobile Multi-Media Wireless Sensor Network (M3WSN) simulation framework. The M3WSN framework has been published as a scientific paper in the 6th International Workshop on OMNeT++ (2013) [1]. M3WSN framework enables the multimedia transmission of real video sequence. Therefore, a set of multimedia algorithms, protocols, and services can be evaluated by using QoE metrics. Moreover, key video-related information, such as frame types, GoP length and intra-frame dependency can be used for creating new assessment and optimization solutions. To support mobility, M3WSN utilizes different mobility traces to enable the understanding of how the network behaves under mobile situations. This tutorial will cover how to install and configure the M3WSN framework, setting and running the experiments, creating mobility and video traces, and how to evaluate the performance of different protocols. The tutorial will be given in an environment of Ubuntu 12.04 LTS and OMNeT++ 4.2.

## I. Introduction

The rapid development of low-cost technologies involving camera sensors and scalar sensors have made Wireless Multimedia Sensor Networks (WMSNs) emerging topics. WMSNs promise a wide range of applications in Internet of Things (IoT) and Smart cities, such as environment surveillance, traffic monitoring, etc.

Many OMNeT++ frameworks have been proposed to study protocols in wired and wireless networks, in which Castalia includes advanced wireless channel and radio models, power consumption models, as well as MAC and routing protocols for wireless sensor networks (WSNs). However, Castalia does not provide any functionality for video transmission, control and evaluation as expected for emerging multimedia applications. This is mainly due to the fact Castalia was designed for scalar sensor network simulation.

Wireless Simulation Environment for Multimedia Networks (WiSE-MNet) [2] incorporates some of Castalias functionalities/features to provide a generic network-oriented simulation environment for WMSNs. WiSE-MNet addresses the need for co-designing network protocols and distributed algorithms for WMSNs. Even though designed for WMSNs, WiSE-MNet does not provide video control and QoE support, which is a key characteristic to enable multimedia evaluation from the users perspective. Additionally, it considers an idealistic communication mechanism to test algorithms without taking into account the unreliable nature of wireless medium. Moreover, WiSE-MNet does not support node mobility with complex traces as expected in many smart cities applications.

The Wireless Video Sensor Network (WVSN) model proposes a simulation model for video sensor networks. It defines the sensing range of camera nodes by a Field of View (FoV), which is more realistic for WMSNs. Additionally, depending on the number of nodes, the model determines the cover-sets for each sensor node and computes the percentage of coverage for each cover-set. Then, this information is used to increase the frame capture rate, e.g., a node has a higher capture rate when it has more covers. However, this work also fails in providing an accuracy video transmission and evaluation approach, and no mobility is supported.

In this context, it is clear that the existing OMNeT++ frameworks have no support for transmission, control and evaluation of real video sequences as required for many WMSNs and smart cities scenarios. Therefore, a QoE-aware and video-related framework that manages video flows with different characteristics, types, GoP lengths, and coding, are required. This framework should also be able to collect information about the type of every received/lost frame, frame delay, jitter and decoding errors, as well as inter and intra-frame dependency of the received/distorted videos, such that a set of mobile multimedia-based protocols can be evaluated and improved. M3WSN extends Castalia by integrating the functionalities of both WiSE-MNet and WVSN models, such that it supports transmission, control and evaluation of real video sequences in mobile WMSNs.

## II. Getting Started

To support mobile object detection and tracking, certain image/video processing libraries are required. The Open Source Computer Vision Library (OpenCV) [3] is the most used library to detect, track, and understand the surrounding world captured by image sensors. OpenCV is released under a BSD license and hence it is free for both academic and commercial use. It has C++, C, Python and Java interfaces and supports different operating systems. The library can take advantage of multi-core processing. Enabled with OpenCL, it can take advantage of the hardware acceleration of the underlying heterogeneous compute platform. Therefore, the first step is to install and configure OpenCV library.

## III. OpenCV Installation

### A. Installation Steps

To install and configure OpenCV, complete the following steps. The commands shown in each step can be copied and pasted directly into a Linux command line.

1. Remove any installed version of ffmpeg and x264:

```
$ sudo apt-get remove ffmpeg x264 libx264-dev
```

2. Get all the dependencies for x264 and ffmpeg:

```
$ sudo apt-get update
$ sudo apt-get install build-essential
    checkinstall git cmake libfaac-dev libjack-
    jackd2-dev libmp3lam-dev libopencore-amrnb-dev
     libopencore-amrwb-dev libsdl1.2-dev libtheora
    -dev libva-dev libvdpau-dev libvorbis-dev
    libx11-dev libsfixes-dev libxvidcore-dev
    texi2html yasm zlib1g-dev
```

3. Download and install gstreamer:

```
$ sudo apt-get install libgstreamer0.10-0
    libgstreamer0.10-dev gstreamer0.10-tools
    gstreamer0.10-plugins-base libgstreamer-
    plugins-base0.10-dev gstreamer0.10-plugins-
    good gstreamer0.10-plugins-ugly gstreamer0.10-
    plugins-bad gstreamer0.10-ffmpeg
```

4. Download and install gtk:

```
$ sudo apt-get install libgtk2.0-0 libgtk2.0-dev
```

5. Download and install libjpeg:

```
$ sudo apt-get install libjpeg8 libjpeg8-dev
```

6. Download, install, configure, and build x264 libraries:

```
$ wget ftp://ftp.videolan.org/pub/videolan/x264/
    snapshots/x264-snapshot-20120528-2245-stable.
    tar.bz2
$ tar xvf x264-snapshot-20120528-2245-stable.tar.
    bz2
$ cd x264-snapshot-20120528-2245-stable.tar.bz2
$ ./configure --enable-static
$ make
$ sudo make install
```

7. Download, install, and configure ffmpeg libraries:

```
$ wget http://ffmpeg.org/releases/ffmpeg-0.11.1.
    tar.bz2
$ tar xvf ffmpeg-0.11.1.tar.bz2
$ cd ffmpeg-0.11.1
$ ./configure --enable-gpl --enable-libfaac --
    enable-libmp3lame --enable-libopencore-amrnb
    --enable-libtheora --enable-libvorbis
$ make
$ sudo make install
```

8. Download and install v4l (video for Linux):

```
$ wget http://www.linuxtv.org/downloads/v4l-utils/
    v4l-utils-0.8.8.tar.bz2
$ tar xvf v4l-utils-0.8.8.tar.bz2
$ cd v4l-utils-0.8.8
$ make
$ sudo make install
```

9. Download and install OpenCV libraries:

```
$ wget http://downloads.sourceforge.net/project/
    opencvlibrary/opencv-unix/2.4.2/OpenCV-2.4.2.
    tar.bz2
$ tar xvf OpenCV-2.4.2.tar.bz2
$ cd OpenCV-2.4.2
$ make build
$ cd build
$ cmake -D CMAKE_BUILD_TYPE=RELEASE
```

10. Configure Linux:

```
$ export LD_LIBRARY_PATH=/usr/local/lib
$ PKG_CONFIG_PATH=$PKG_CONFIG_PATH:/usr/local/lib/
    pkgconfig
$ export PKG_CONFIG_PATH
```

### B. Possible Problems

During the installation process of OpenCV, several problems might be encountered. In this tutorial, we will discuss the possible errors that might block the installation procedure, and their corresponding solutions.

## IV. M3WSN Framework Installation

### A. Downloading M3WSN

The M3WSN framework can be downloaded from http://cds.unibe.ch/research/M3WSN/index.html. After downloading, unzip the file, which includes a M3WSN folder and a customized version of Castalia.

### B. Building the Project

1. Import Castalia-3.2 M3WSN and M3WSN project to OMNeT++ IDE.

2. Check if M3WSN has reference to Castalia-3.2 M3WSN.

3. Clean and build the project.

## V. Creating Video Sequences

Applications involving multimedia transmission must be evaluated by measuring the video quality level from the user's perspective. Due to the importance of the multimedia content, it is essential to visually determine the real impact of the event, perform object/intruder detection, and analyze the scenes based on the collected visual information. Specifically, frames with different priorities (I, P and B) compose a compressed video, and the loss of high priority frames causes severe video distortion from humans experience. For the loss of an I-frame, the errors propagate through the rest of the Group of Picture (GoP), because the decoder uses the I-frame as the reference frame for other frames within a GoP. However, Castalia framework, including both WiSE-MNet and WVSN models, does not enable the transmission, controlling and evaluation of real video sequences. To this end, we ported Evalvid [3] for the M3WSN framework. Evalvid provides video-related information, such as frame types, received/lost frames, delay, jitter, and decoding errors, as well as inter and intra-frame dependency of the received/distorted videos. These information will be helpful to design new video transmission protocols.

Evalvid is a framework for video transmission and quality evaluation. Therefore, before transmitting a real video sequence, we need a video source, for instance from a video library or the user can create a new one. Once the video has been encoded, trace files have to be produced. The trace files contain all the relevant information for video transmission, and the evaluation tools provide routines to read and write these traces files for multimedia evaluation. Information about how to create video traces using Evalvid can be found in http://www2.tkn.tu-berlin.de/research/evalvid/EvalVid/docevalvid.html.

## VI. Creating Mobility Traces

Mobility is one of the most challenging issues in WMSNs. To understand how the network behaves under different mobility situations, the node mobility has to be simulated in a reasonable way. In this context, we have ported BonnMotion to M3WSN to support mobility. BonnMotion is a simulator-independent tool to generate mobility traces for different mobility models. It provides several mobility models, such as the Random Waypoint model, the Gauss-Markov model, and others. The generated mobility traces can be exported to compatible simulator. Information about how to create mobility traces can be found at http://sys.cs.uos.de/bonnmotion/.

## VII. Running Experiments

Experiments can be run using both simulation IDE or using command line. We show two approaches in below.

### A. Using OMNeT++ IDE

After successfully importing and building the project, simulations can be run by following the next steps:
1. Open some configuration files (ini file) from M3WSN/Simulations folder to run the simulations. The configuration file (ini file) describes the experiment scenario and it includes all the settings of the involved modules, i.e., applied protocols at different network layers, protocol parameters, simulation duration, etc. For instance, routing protocol/parameters, application module/parameters, and radio module/parameters can be configured as below:

```
$SN.node[*].Communication.RoutingProtocolName = "
    GPSR"
$SN.node[*].Communication.Routing.netBufferSize =
    64
$SN.node[*].Communication.Routing.
    netDataFrameOverhead = 6

$SN.node[*].Communication.Radio.TxOutputPower = ${
    TxPower="-5dBm", "-10dBm", "-15dBm"}
$SN.node[0].Communication.Radio.CCAthreshold = ${
    CCAthreshold=-95, -90, -85}

$SN.node[*].ApplicationName = "ThroughputTest"
$SN.node[*].Application.packet_rate = 5
$SN.node[*].Application.constantDataPayload = 2000
```

2. Select "Command line" in the "Options" before running the experiments, as shown in Figure 1.

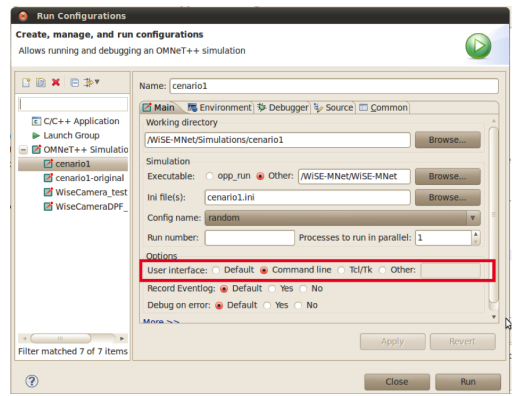

Fig. 1. Simulation configuration using IDE

### B. Using command line

In addition to running the experiment using OMNeT++ IDE, it is also possible to run simulation using command line. After building the project using IDE, the scripts at M3WSN/Simulations folder can be used to start the experiments. Additionally, M3WSN framework provides scripts to reconstruct video sequence, measure video quality level, and run simulations with different random-generated seeds.

## VIII. Simulation Outputs

After the simulation is finished, results will be saved in different files.

### A. M3WSN-result.txt

The file M3WSN-file.txt is automatically generated as done in Castalia. It contains a summary of simulation procedure. The user can use some scripts to filer the results, which can be found in section 3.3 of Castalia user manual.

This file is used to generate customized outputs by using "output()" function to the *.cc* file. Then, the user can use some external tool to analyze the file, and extract some results from them. By default, all tracing is turned off, and the user can turn on for each module in the *.ini* file. For example:

```
$ SN.node[*].ResourceManager.collectTraceInfo =
    True
$ SN.node[*].SensorManager.collectTraceInfo =
    False
$ SN.node[*].Communication.Routing.
    collectTraceInfo = True
$ SN.node[*].Communication.MAC.collectTraceInfo =
    True
$ SN.node[*].Communication.Radio.collectTraceInfo
    = True
$ SN.wirelessChannel.collectTraceInfo = True
```

An example of M3WSN-result.txt is shown as below:

```
Transmitted Videos
Time       Video-id      Node       Hops
4.13594    0             1          5
24.1421    1             1          7
44.1471    2             1          6
64.1513    3             1          6
84.2111    4             1          6
104.154    5             1          6
```

### B. M3WSN-Debug.txt

It contains a trace of all events that the user has requested to be recored by "turning on" some parameters in the *.ini* file. This file can be used to debug the code. The user has to use the command "trace()" in the *.cc* file to add information into this file. By default, all tracing is turned off, and the user can turn on for each module in the *.ini* file as for the M3WSN-result.txt

Here we also give show an example of M3WSN-Debug.txt:

```
0.027540267327 SN.node[1].Application Node 1 is
    sending packets
3.868523146136 SN.node[0].Application Received
    packet #18 from Node 1
4.062451653241 SN.node[0].Application Received
    packet #19 from Node 1
4.263512358156 SN.node[0].Application Received
    packet #20 from Node 1
4.463984107516 SN.node[0].Application Received
    packet #21 from Node 1
4.668401238515 SN.node[0].Application Received
    packet #22 from Node 1
```

From the generated output files, we could use some scripts to extract the information of interest, such as to calculate the packet delivery ratio of a node by checking the number of transmitted packets and the number of packets successfully received by other nodes.

## IX. Performance Evaluation

In this section, we introduce a use case that makes use of M3WSN framework to obtain key video-related information, such as frame type and GoP length for creating new assessment and optimization solutions. Additionally, the described use case shows the importance to evaluate the transmitted video sequences from the user's perspective. This use case scenario can be easily extended to smart cities applications.

### A. Scenario Description

Multimedia video transmission is applicable to many situations, such as multimedia surveillance, real-time traffic monitoring, personal health care, and environmental monitoring. In this tutorial, we apply M3WSN framework in an intrusion detection scenario where static scalar and camera sensors are deployed to monitor a corridor to detect any intruder. The intrusion detection approach is based on our proposed QoE-aware FEC (Forward Error Correction) mechanism for intrusion detection in multi-tier WMSNs, which was described in [4]. In the scenario, a set of sensors performs intrusion detection using vibration sensors. Another set of camera nodes only transmit real-time videos from the intruder area, once the scalar sensors detected it. At the camera nodes, the QoE-aware FEC mechanism creates redundant packets based on frame importance from user's experience, and thus reduce the packet overhead, while keeping the video with a good quality.

On the basis of the multi-tier intrusion detection scenario, simulation can be carried out to evaluate the transmitted video from user's perspective by using QoE-aware FEC. Following this, we simulated a simple FEC approach, i.e., creating redundancy for all the frames (simple FEC), and also without any FEC mechanism (no-FEC). The simulations were carried out and repeated 20 times with different random seed numbers to provide a confidence interval of 95%. Table I shows the simulation parameters for these solutions.

| Parameter | Value |
| --- | --- |
| Field Size | 80x80 |
| Location of Base Station | 40, 0 |
| Initial location of intruder | 0, 0 |
| Intruder movement type | Random mobility |
| Intruder velocity | 1.5 |
| Total number of Nodes | 100 |
| Number of nodes at high-tier | 25 |
| High-tier deployment | Grid |
| Low-tier deployment | Uniform |
| Transmission Power | -15 dbm |
| Path loss model | Lognormal shadowing model |
| Radio model | CC2420 |
| Video sequence | Hall |
| Video Encoding | H.264 |
| Video Format | QCIF (176 x 144) |
| Frame Rate | 26 fps |

The intruder starts at location (0,0), and moves in a random way. The low-tier nodes have an ominidirectional sensing range, and detect the intruder by using the intruder bounding boxes. As soon as the low-tier detects the intruder, it must wake up the high-tier to send the video of the detected intruder. Video flows provide more precise information for users and authorities (e.g. the police) about the intruder, and enable them to monitor, detect, and predict the intruder's moving direction. Additionally, they allow the authorities to take precise actions in accordance with the visual information.

### B. Performance Metrics

Existing works on multimedia area classify the videos into three categories, according to their motion and complexity, i.e. low, median and high. For example, Aguiar et al. classify the Hall video sequence (taken from the Video Trace Library) as low movement, which means that there is a small moving region on a static background, i.e. men walking in a hall [?].

We evaluated the transmitted videos by means of two well-known objective QoE metrics, i.e. Structural Similarity (SSIM) and Video Quality Metric (VQM), obtained by using the MSU Video Quality Measurement Tool (VQMT) [5]. SSIM measures the structural distortion of the video, and attempts to obtain a better correlation with the user's subjective impression. SSIM has values ranging from 0 to 1, a higher value meaning a better video quality. On the other hand, VQM measures the "perception damage" of video experienced, based on features of the human visual system, including distinct metric factors such as blurring, noise, color distortion and distortion blocks. A VQM value closer to 0 means a video with a better quality.

### C. Simulation Results

In the experiments, we measure the SSIM and VQM for transmitted videos with respect to the length of the transmission route (hop numbers), as shown in Figure 2 and Figure 3

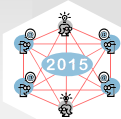

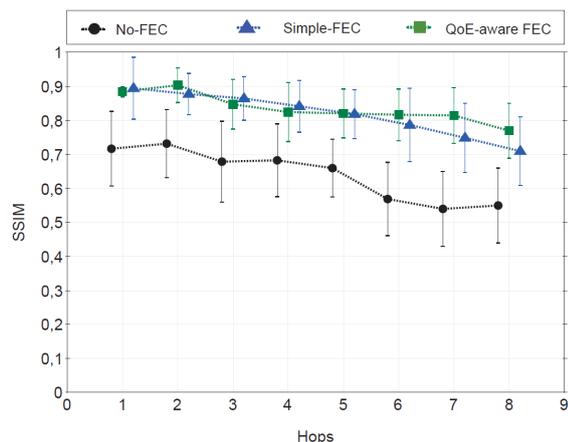

Fig. 2. SSIM with respect to number of hops

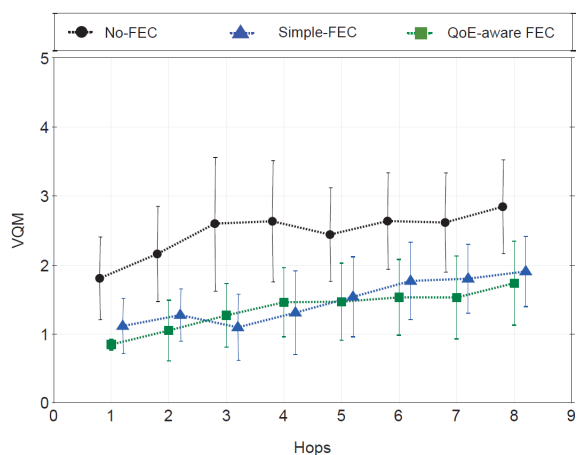

Fig. 3. VQM with respect to number of hops

Figure 2 shows that solutions create redundant packet improve the SSIM by around 25% compared to solutions without FEC. This is due to the fact that application-level FEC is applied as error control scheme for handling packet losses in real-time transmissions. Hence, the redundant packets can be used to reconstruct a lost frame, and thus improve the video quality from a user's perspective. Due to less transmission means less energy consumption, we can conclude that QoE-aware FEC can provide energy-efficiency, while keeping the transmitted video with a good quality level. Tis is because QoE-aware FEC creates redundant packets based on frame importance and user experience to reduce network overhead.

Figure 3 presents the video quality by using VQM. The VQM results demonstrate the benefits of using FEC and confirm the SSIM values. Both simple and QoE-aware approaches kept the VQM vales below the solution without FEC, and thus improve the video quality level. However, the QOE-aware FEC mechanism reduces the amount of generated redundant packet while keeping videos with an acceptable quality level.

Last, to show the impact of transmitting video streams from the standpoint of an end-user, a frame was randomly selected (Frame 258) from the transmitted video, as displayed in Figure 4. Frame 258 is the moment when a man (the intruder in our application) was walking along a corridor. For intruder detection application, this is an important frame to provide users and authorities with more precise information and allow them to make actions. The benefits of the FEC mechanisms are visible by analyzing the frames in Figure 4. By comparing each transmitted frame with the original one, it is possible to see a higher distortion for the frame transmitted without using any FEC, as shown in 4(a). The frames transmitted using FEC mechanism achieves low distortion, as shown in 4(c) and 4(d). The visual evaluation is only possible due to M3WSN supports the transmission and control of real video sequences.

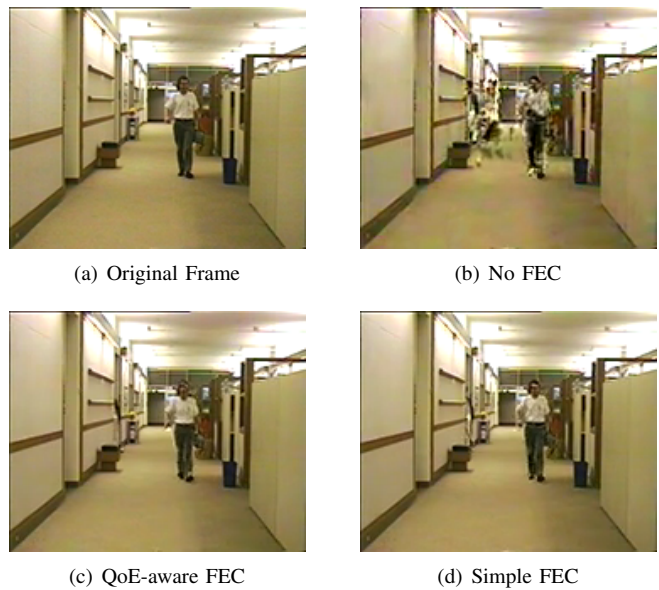

(a) Original Frame      (b) No FEC

(c) QoE-aware FEC      (d) Simple FEC

Fig. 4. Frame 258 of transmitted video

## X. CONCLUSION

This tutorial gives detailed explanations about how to install and configure the M3WSN OMNeT++ framework, which enables the transmission of real video sequence. M3WSN framework supports mobility, and it can generate real video sequences. During the tutorial, we also discuss the possible problems that might be encountered during the installation process and the corresponding solutions.

M3WSN can also be used to evaluate protocols at different network stacks, e.g., routing protocols, transport protocols, or audio/video codes mechanisms.